\documentclass{elsart5p}

\journal{Diamond and Related Materials}

\usepackage{graphicx}
\usepackage{amssymb}
\usepackage[pagewise,switch]{lineno}
\begin{document}

\begin{frontmatter}
\title{Structure, elastic properties and strength of amorphous and nanocomposite carbon}
\author[physics]{Ioannis N. Remediakis\corauthref{cor1}}
\ead{remed@physics.uoc.gr}
\author[physics]{Maria G. Fyta\thanksref{Harvard}}
\author[materials]{Christos Mathioudakis}
\author[materials]{Georgios Kopidakis}
\author[physics,forth]{Pantelis C. Kelires}
\address[physics]{Department of Physics, University Of Crete, P.O. Box
2208, 71003 Heraklion, Crete, Greece}
\address[materials]{Department of Materials Science and Technology,
University of Crete, P.O. Box 2208, 71003 Heraklion, Crete, Greece }
\address[forth]{IESL, Foundation for Research and Technology-Hellas
(FORTH), 71003 Heraklion, Crete, Greece}

\corauth[cor1]{Corresponding author. Present address: Department of Materials
Science and Technology, University of Crete, P.O. Box 2208, 71003 Heraklion,
Crete, Greece.}
\thanks[Harvard]{Present address: Department of Physics, Harvard University,
  Cambridge, MA 02138, USA}

\begin{abstract}
We study theoretically the equilibrium structure, as well as the response
under external load, of characteristic carbon-based materials. The materials
considered include diamond, amorphous carbon (a-C), ``amorphous diamond'' and
nanocomposite amorphous carbon (na-C). A universal bulk-modulus versus density
curve is obeyed by all structures we consider. We calculate the dependence of
elastic constants on the density. The strength of a-C was found to increase in
roughly a linear manner, with increasing concentration of four-fold atoms,
with the maximum stress of the strongest a-C sample being about half that of
diamond. The response of na-C to external load is essentially identical to the
response of the embedding a-C matrix.
\end{abstract}

\end{frontmatter}


\section{Introduction}
\label{sec:intro}

Amorphous carbon (a-C) is a well-established carbon-based material, with
distinct mechanical properties such as hardness and thermal stability. These
properties render a-C to be an ideal material for coating
applications. Recently, nanocomposite carbon (na-C) has been synthesized in
both hydrogenated and pure forms \cite{lifshitz02yao05}. It is considered as a
material with similar, and perhaps better properties than those of a-C. Na-C
consists of nanometer-scale regions of crystalline material embedded into an
a-C matrix. This material has mechanical properties between those of a-C and
diamond \cite{fyta06}, that could in principle be fine-tuned by adjusting
either the size of the nanocrystalline region or the density of the amorphous
matrix. Such tailoring of material properties to match those desired in the
applications could be invaluable.

The properties of a-C and na-C can be categorized into two broad classes: The
first class includes macroscopic, mean-field properties, such as the density,
that are implicitly only related to individual chemical bonds.  Atomistic
properties, on the other hand, such as the strength, arise directly from the
individual chemical bonds between C atoms of different electronic
structure. The macroscopic properties are usually related to the minimum or
the harmonic regime of the cohesive energy versus volume curve, and can,
therefore, be easily calculated using any reasonable description for the
bonds. On the other hand, even a rough estimate of atomistic properties
requires employing some level of quantum theory.

A quantum-mechanical treatment is able to accurately describe the chemical
bond, but can only be used to study few hundreds of atoms for very short time
scales. On the other hand, a classical empirical potential method can handle
three orders of magnitude more atoms for realistic time scales, at the price
of a limited accuracy in the description of the chemistry of carbon. This
interplay between classical and quantum-mechanical simulations has been used
to verify or predict several experimental findings for na-C: Diamond
nanocrystallites were found to be stable, having negative formation energy,
when the amorphous matrix is dense \cite{fyta03}. The average intrinsic stress
of the material is zero \cite{fyta05}. Although the bulk modulus of na-C is
higher than that of a-C, the two materials have identical ideal strength due
to identical fracture mechanisms \cite{fyta06}. In this work, we first
establish the validity of our simulations by ensuring that different
approximations yield identical qualitative and similar quantitative
results. We then study in more detail the elastic properties of these
materials, including the dependence of elastic moduli as a function of their
density. Finally, we comment on their behavior under strain well beyond the
elastic regime, and discuss briefly their fracture.

\section{Computational Method}
\label{sec:method}

The quantum-mechanical calculations are based on two different tight-binding
hamiltonians. The so-called NRL-TB was developed by Papaconstantopoulos, Mehl
and co-workers at the Naval Research Laboratory \cite{cohen94mehl96}. The
parametrization of the hamiltonian for C is based on similar assumptions to
the previously published parametrization for Si; see Ref. \cite{papacon03} for
a review. The environment-dependent tight-binding (EDTB) model of Wang, Ho and
co-workers \cite{tang96} goes beyond the traditional two-center approximation
and allows the TB parameters to change according to the bonding
environment. In this respect, it is a considerable improvement over the
previous two-center model of Xu {\it et al.} \cite{xu92}. Both NRL-TB and EDTB
schemes have been used successfully to simulate a-C systems
\cite{mathioudakis04,fyta06}. The tight-binding molecular-dynamics simulations
simulations are carried out in the canonical (N, V, T) ensemble, T being
controlled via a stochastic temperature control algorithm. The supercells used
in the tight-binding simulations contain 512 C atoms each.

The empirical potential simulations are based on the continuous-space Monte
Carlo method. We employ the many-body potential of Tersoff \cite{tersoff88},
which provides a very good description of structure and energetics for a wide
range of carbon-based materials \cite{kelires94,kelires00}. This method allows
for great statistical accuracy, as it is possible to have samples at full
thermodynamic equilibrium. Moreover, through the use of relatively large
supercells, it offers the possibility to explore a larger portion of the
configurational space of the problem. The supercells used in the Monte Carlo
simulations contain 4096 C atoms each.

Na-C is modeled by a periodic repetition of cubic supercells that consist of a
spherical crystalline region surrounded by a-C. To construct such supercells,
we first consider a cubic supercell of diamond, and choose the radius of the
spherical crystalline phase. Keeping the atoms inside this sphere frozen to
their diamond lattice positions, we run the system at a very high temperature
(12000-15000 K) so that a liquid is created, and then quench it down to 50 K,
at constant volume. After that, the system is fully equilibrated by relaxing
both the system volume and the coordinates of all atoms. In the Monte Carlo
simulations, we perform an additional intermediate relaxation at room
temperature to ensure that the sample is fully relaxed.

By adjusting the volume or, equivalently, the pressure of the liquid phase, we
can create samples having low- or high-density a-C. It has been shown that
properties of a-C can be described in terms of a single parameter, $z$
\cite{mathioudakis04}. This is the average coordination number, or number of
neighbors, for each atom in the sample. More precisely, $z$ it is the integral
of the nearest-neighbor peak of the pair-correlation function. By convention,
individual atoms in a-C are thought of having an electronic structure in
one-to-one correspondence to their coordination number: four-, three- and
two-fold atoms are usually labeled in the literature as $sp^3$, $sp^2$ and
$sp^1$, respectively. Although for most atoms such a relationship between
coordination number and hybridization holds to a good approximation, other
atomic electronic structures may be present as well. We create samples with
average coordination between 3.1 and 3.9, having concentrations of four-fold
coordinated atoms between 10 and 90\%. Samples with $z \gtrsim $ 3.8 are
considered as tetrahedral amorphous Carbon (ta-C). We also consider the fully
tetrahedral Wooten-Winer-Weaire (WWW) structure \cite{wooten85}, as a model
for ``amorphous diamond''. The radius of the nanocrystalline region is of the
order of 2 nm (1 nm for the quantum-mechanical simulations), while it occupies
about 30-40\% of the total volume, in accordance to experimental observations
\cite{zhou02}.

\section{Equilibrium structural properties}
\label{sec:structure}

\begin{table}
\caption{Structural properties of diamond (D), ``amorphous diamond'' (WWW) and
four characteristic a-C samples (A-D). All samples are simulated using NRL-TB,
except D (E), D (T) and D (e) that contain results obtained from the EDTB,
Tersoff potential and experiment, respectively.  $N_4$, $N_3$ and $N_2$ is the
percentage of four-, three- and two-fold atoms, respectively. $B$ is the bulk
modulus (in GPa), $\rho$ the calculated density (in gr/cm$^3$), and
$\rho^{fit}$ is the density (in gr/cm$^3$) according to the fit of $\rho$
vs. $z$ from Ref. \cite{mathioudakis04}.}
\begin{center}
\begin{tabular}{p{0.11\columnwidth}
                p{0.11\columnwidth}
                p{0.11\columnwidth}
		p{0.11\columnwidth}
		p{0.11\columnwidth}
		p{0.11\columnwidth}
		p{0.11\columnwidth}
		p{0.11\columnwidth}} \hline \hline
        & $z$ & $N_4$    & $N_3$ & $N_2$     & $B$ & $\rho$ & $\rho^{fit}$ \\ \hline
D       & 4.0 &  100     &   0        &    0 & 480 &   3.7      &              \\
D (E)   & 4.0 &  100     &   0        &    0 & 428 &   3.5      &              \\
D (T)   & 4.0 &  100     &   0        &    0 & 422 &   3.5      &              \\
D (e)   & 4.0 &  100     &   0        &    0 & 443 &   3.5      &              \\
WWW     & 4.0 &  100     &   0        &    0 & 434 &   3.4      &    3.5       \\
A       & 3.8 &  78      &   22       &    0 & 387 &   3.1      &    3.2       \\
B       & 3.7 &  67      & 32         &    1 & 361 &   2.9      &    3.0       \\
C       & 3.5 &  47      & 52         & 1    & 325 &   2.5      &    2.7       \\
D       & 3.2 &  18      & 77         & 5    & 211 &   1.6      &    2.1       \\ 
\hline \hline
\end{tabular}
\end{center}
\label{tab:nrl}
\end{table}

Structural properties of diamond, ``amorphous diamond'' and a-C are summarized
in Table \ref{tab:nrl}. For diamond, we calculate the density ($\rho$) and the
bulk modulus ($B$) using the two tight-binding schemes and the Tersoff
potential.  All methods yield results in good agreement to experiment. The a-C
samples we considered consist mostly of four- and three-fold atoms, while
two-fold atoms appear in low-density samples. Using the EDTB model, we had
found that the density of a-C depends linearly on $z$, via $\rho\approx
-3.3+1.7z$ \cite{mathioudakis04}. We find that the linear relationship found
using EDTB is also valid within NRL-TB, at least for the denser samples.

\begin{figure} 
\begin{center}
\includegraphics[width=\columnwidth]{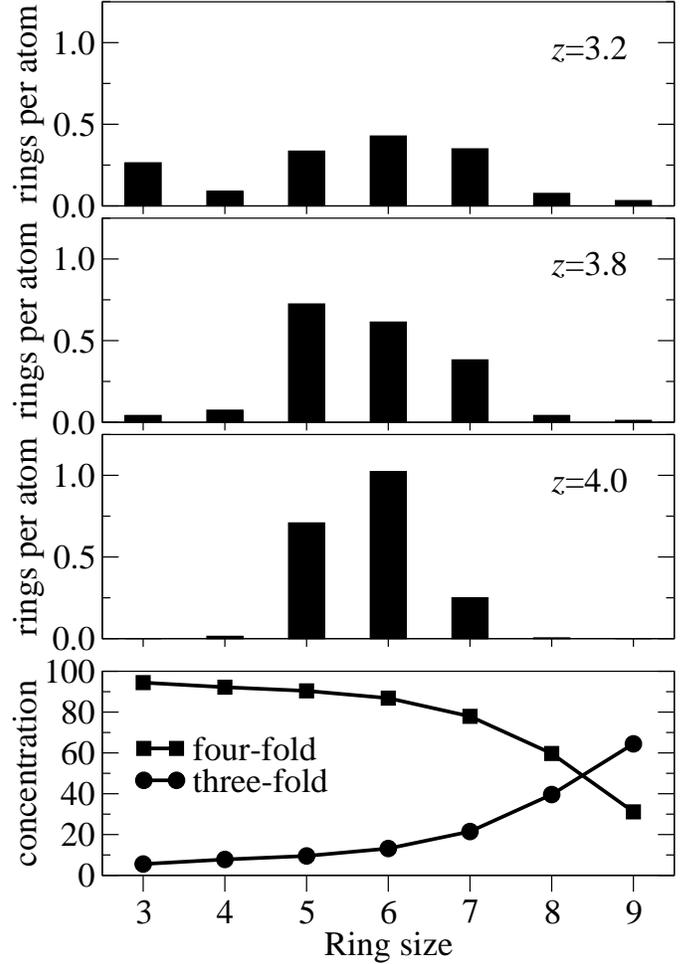} 
\end{center}
\caption{Ring statistics for various a-C samples, as calculated using the
  NRL-TB. The upper three panels show the number of rings divided by the total
  number of atoms in the sample, as a function of the number of atoms
  participating in each ring, for three cases: low-density a-C ($z$=3.18),
  ta-C (z=3.78) and ``amorphous diamond'' (z=4.00). Lowest panel shows the
  concentration of rings in four-fold and three-fold atoms for the ta-C
  sample.}
\label{fig:rings}
\end{figure}

Insight into the atomistic structure of a-C can be gained by looking at the
sizes and distributions of rings of atoms in the material. Rings of atoms are
defined in terms of the shortest-path criterion of Franzblau
\cite{franzblau91}. The number of atoms that participate in a ring is the ring
size, and can have only specific values for a bulk crystalline material: in
diamond, for example, the smallest ring size is six. On the contrary,
three-member rigs are common in a-C. As the formation of rings is related to
atomistic binding, a quantum-mechanical treatment is necessary in order to get
reliable ring statistics.

Results for the relative ring numbers and concentrations, found using the
NRL-TB for three characteristic samples, are shown in
Fig. \ref{fig:rings}. Similar results have been obtained with EDTB
\cite{mathioudakis04}. For the low-density sample, we find significant numbers
for three- and four- member rings, while the most probable ring length is
six. For ta-C, the most probable ring length is five, while there are still
some three- and four- member rings. Finally, for the ``amorphous diamond''
sample, most rings are six-membered, with few five- and seven-member
ones. Although, by construction, only four-fold atoms exist in this sample,
its random topology allows for rings that would not be present in the
crystalline material.

The bottom panel of Fig. \ref{fig:rings} shows the composition of rings as a
function of their size, for the ta-C sample. Most atoms (over 90\%) of the
smallest rings are four-fold. The concentration of four-fold atoms decreases
with increasing ring size; for large rings, with more that eight members, most
atoms are three-fold. This behavior can be attributed to the long-range
correlations found for $\pi$-bonded atoms; such $\pi$ bonds are more likely to
occur between three- than between four-fold atoms. On the other hand, such
double bonds are more difficult to bend in order to form triangles or
quadrangles; this is why most atoms that participate in small rings are
four-fold. 85 \% of the three-member rings and 75 \% of the four-member rings
consist of four-fold atoms only, while there are no rings of sizes 8 and above
having only four-fold atoms. The ring statistics presented here are in
excellent agreement to experiments and state-of-the-art {\em ab initio}
simulations \cite{mcculloch00,marks02}.

\begin{figure} 
\begin{center}
\includegraphics[width=\columnwidth]{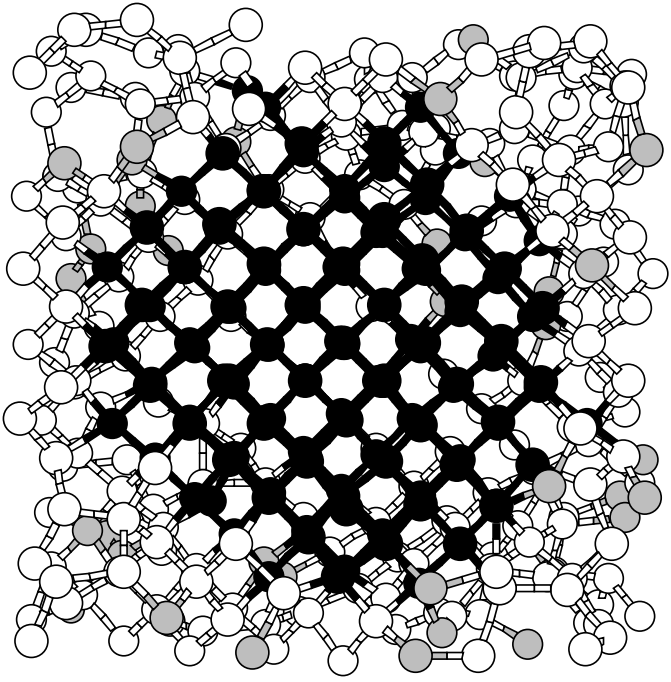} 
\end{center}
\caption{Cross-section of a ball-and-stick representation for a typical model
structure of nanocomposite amorphous Carbon created using EDTB: a spherical
nanocrystal with a diameter of 1.24 nm is embedded into amorphous-C with an
average coordination of 3.8. Atoms belonging to the crystal are represented by
black spheres; four-and three-fold atoms of the amorphous matrix are
represented by white and gray spheres, respectively.}
\label{fig:structure}
\end{figure}

A typical sample of na-C, generated in the NRL-TB scheme is shown in
Fig. \ref{fig:structure}. The volume fraction of the nanocrystalline region is
31\%. The surrounding amorphous material is ta-C ($z$=3.8 and $\rho$=3.0
gr/cm$^3$). A sample created using EDTB under identical conditions has the
same crystal volume fraction and a-C coordination, and only slightly higher
density of the amorphous phase, 3.1 gr/cm$^3$. In all cases, the surrounding
a-C matrix obeys the same density vs. coordination relationship as pure a-C
samples generated using the same recipe. The a-C atoms are covalently bonded
to the crystal, resulting thus in thin interface regions.

\begin{figure} 
\begin{center}
\includegraphics[width=\columnwidth]{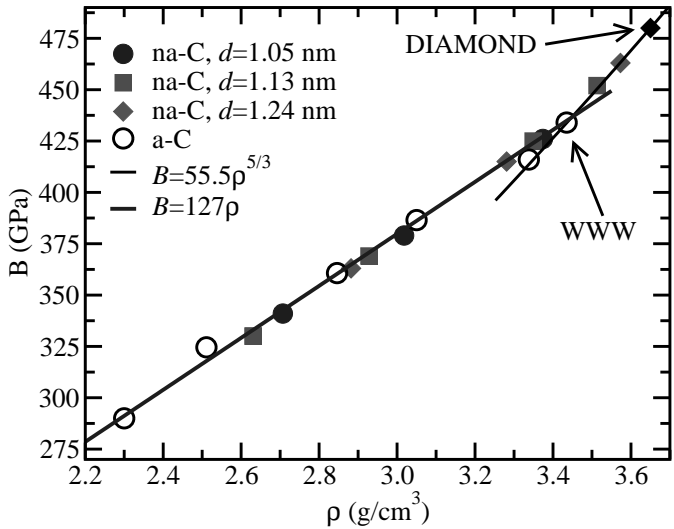} 
\end{center}
\caption{Bulk modulus as a function of density, calculated within NRL-TB, for
  na-C with different radii of the crystalline region. Calculations for a-C of
  various densities, bulk diamond and ``amorphous diamond'' are also shown.}
\label{fig:bulkmodulus}
\end{figure}

As a first estimate of the hardness of nanocomposite nd/a-C structures, we
calculated their bulk modulus, $B$. In Fig. \ref{fig:bulkmodulus}, we plot $B$
as a function of the total density of the samples, $\rho$. The bulk modulus of
na-C is enhanced compared to that of pure a-C \cite{fyta06}. Replacement of
some amorphous material by crystalline increases noticeably the bulk modulus,
rendering it for some samples to be higher even than that of the ``amorphous
diamond'', and close to that of diamond \cite{fyta06}. 

Interestingly, all samples, including pure a-C, na-C, WWW, and even diamond,
seem to follow the same universal curve in Fig. \ref{fig:bulkmodulus}. Such
universalities have been observed in the past: He and Thorpe \cite{he85}
showed that $B\sim (z-z0)^{1.5}$, where $z0$ is universal; Liu {\it et al.}
\cite{liu91} showed that $B\sim d^{-3.5}$, where $d$ is the average distance
between atoms. Recently, Mathioudakis {\it et al.} \cite{mathioudakis04}
showed that the two approaches are equivalent, and that for a-C, $B \sim
(d-\mbox{const.})^{-3.5}$. The last relationship would imply that $B \sim
(\rho^{-1/3}-\mbox{const.})^{-3.5}$. Indeed, this function fits very well the
data of Fig. \ref{fig:bulkmodulus}.

The universal dependence of $B$ on $\rho$ can be understood by considering the
microscopic response of the material to the external pressure. We can think of
two regimes: in low-density materials, the pressure is undertaken by
appropriately adjusting the volume of the void regions of the material. Such
voids exist in every low-density material and are a result of induced dipole
(van der Waals) interactions. On the other hand, for dense materials, strong
covalent bonds have to be deformed, resulting in higher bulk moduli. In this
case, the resistance of the electrons to compression follows from their
quantum nature and the Pauli principle.

In the first case, the scaling of $B$ with respect to $\rho$ can be found by
considering a model solid bonded exclusively through van der Waals
interactions. Using the Lennard-Jones potential we can find that $B \sim \rho$
\cite{ashcroft}. On the other hand, for high densities, we can get the correct
scaling by using the free-electron approximation: in this case, $B \sim
\rho^{5/3}$ \cite{ashcroft}. This picture is demonstrated in
Fig. \ref{fig:bulkmodulus}: $B \sim \rho $ for $\rho \lesssim$ 3.3 gr/cm$^3$,
while $B \sim \rho^{5/3}$ for $\rho \gtrsim $ 3.3 gr/cm$^3$. These
relationships hold with surprisingly good accuracy.

Being such a fundamental average property, the $B$ vs. $\rho$ curve should be
reproduced well by both our tight-binding models and the empirical potential,
as the latter is known to reproduce correctly the elastic response of the
material. On the other hand, the density dependence of elastic constants
associated with changes in shape, like the Young's modulus, can be
different. Although the response of materials to volume changes can be
addressed at almost any level of theory, understanding the response to shape
changes requires a model that takes into account the directionality of the
chemical bonds. Fortunately, as the calculation of elastic constants require
only small deformations from the minimum energy structure, the empirical
potential approach should suffice for their calculation. As the Monte Carlo
simulations offer greater statistical accuracy and refer to more realistic
sizes of the crystalline regions, we prefer to employ this method for the
calculation of elastic constants.

\begin{figure} 
\begin{center}
\includegraphics[width=\columnwidth]{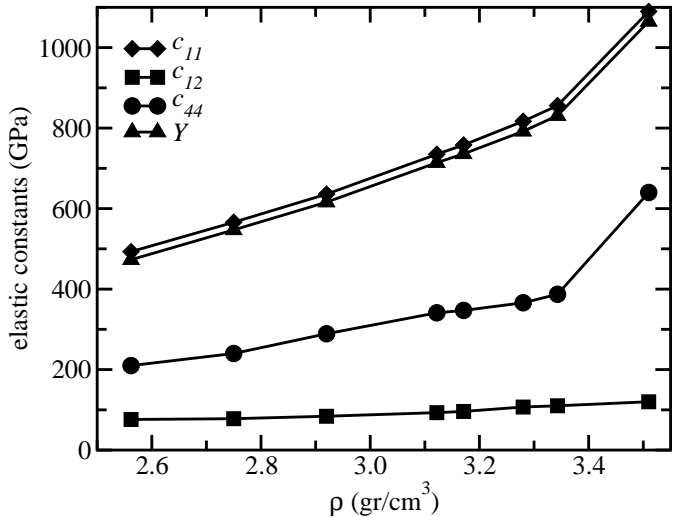}
\end{center}
\caption{Elastic constants of na-C, calculated using the Tersoff potential, as
  a function of the total density. The average diameter of the crystalline
  region is 1.7 nm, and the surrounding a-C matrix has $z$=3.8. The rightmost
  point for each data set, for a density $\rho$ = 3.51 gr/cm$^3$, corresponds
  to diamond and is taken from Ref. \cite{kelires94}.}
\label{fig:elastic}
\end{figure}

To calculate the elastic constants, we apply the appropriate deformation to
the system and compute its total energy as a function of the imposed strain.
The curvature of this function at its minimum yields the desired modulus.  The
number of independent elastic constants depends on the symmetry of the
material: For a material with cubic symmetry, there are three independent
elastic constants, while for an isotropic material, such as a-C or na-C, there
are only two \cite{kaxirasbook}. In Fig. \ref{fig:elastic} we plot $c_{11}$,
$c_{12}$, $c_{44}$ and the Young's modulus $Y$ as a function of density for
na-C and diamond. All elastic constants increase with increasing density,
similar to the previously described behavior of the bulk modulus as a function
of density. For an isotropic material, $2c_{44}=c_{11}-c_{12}$. This
relationship holds within 4\% or less, for all data points presented in
Fig. \ref{fig:elastic}, demonstrating that na-C is a highly isotropic
material. The moduli $c_{11}$ and $c_{44}$ are both associated with changes in
shape, and this is why their values for na-C are much lower than the
corresponding values for diamond, where strong directional bonds are bent.  On
the other hand, $c_{12}$ describes simultaneous elongation along two axes
without shearing, and the value of this modulus for diamond follows the trend
observed for na-C.

\section{Ideal Strength and fracture}
\label{sec:fracture}

The response of covalently-bonded materials, such as a-C and na-C, under
strain can be categorized into three broad regimes: For small strains, the
response of the material is elastic, and Hooke's law stands: the stress is
proportional to the applied strain. For example, if tensile strain is applied
to an isotropic material, the stress will equal the strain times the Young's
modulus. The second regime corresponds to strain beyond the elastic limit, and
is usually associated with plastic deformation of the material. The stress
experienced by the material increases with increasing strain until a maximum
stress (strength) is reached. The third regime is associated with strain
beyond that giving the maximum stress. For brittle materials like diamond, the
material breaks when the maximum stress is reached, and further increase of
the strain results therefore in zero stress. Ductile materials, on the other
hand, can be deformed beyond the strain corresponding to the maximum of the
stress.

\begin{figure} 
\begin{center}
\includegraphics[width=\columnwidth]{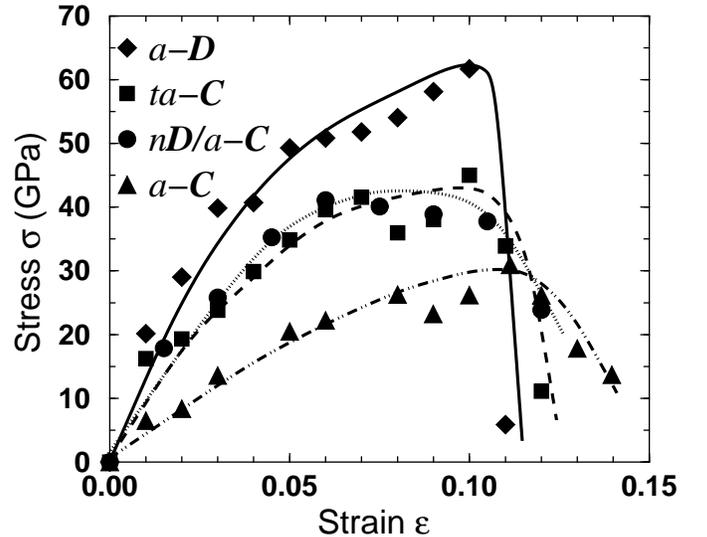} 
\end{center}
\caption{Stress vs. strain curves for WWW model of ``amorphous diamond''
(diamonds), ta-C (squares), low-density a-C (triangles) and na-C
(circles). The latter consists crystalline regions of 1.2 nm surrounded by
ta-C. Data shown are obtained by the NRL-TB method for tensile load in the
(111) direction of the crystal.}
\label{fig:stressstrain}
\end{figure}

To study the strength of a-C and na-C, we apply tensile load on the [111] easy
slip plane of the crystalline region. Strain is simply the ratio of the volume
change divided by the initial volume of the sample; stress is the negative
derivative of the energy with respect to volume. We consider three a-C
samples: a typical a-C sample with average coordination $z$=3.47, a ta-C
sample with $z$=3.8 and an ``amorphous diamond'' sample.  As graphite ($z$=3)
is known to be much softer than diamond ($z$=4), it is reasonable to expect
that the strength increases with increasing $z$, as was the case for the
elastic constants discussed in Section \ref{sec:structure}. This is observed
in Fig. \ref{fig:stressstrain}. The maximum stresses are roughly 60, 40 and 30
GPa for $z$=4.00, 3.78 and 3.50, respectively, so that the strength of a-C is
roughly proportional to its concentration of four-fold atoms. Interestingly,
the stress versus strain curve for the low-density a-C sample seems to suggest
a ductile behavior. We get similar results when applying shear strain. As a-C
is highly isotropic, the energy required to deform the material is a function
of the change in its volume and does not depend much on how this change is
applied.

The strength of diamond does not fit into the simple picture of the strength
being proportional to the concentration of four-fold atoms. If that was the
case, then the strength of the isotropic WWW sample under tensile load would
be higher than the strength of diamond under tensile load perpendicular to its
easy slip plane, (111). After all, in the WWW model the number of bonds per
unit area for a given direction has to be higher than the number of bonds per
unit area on the [111] plane of diamond. This justifies the name ``easy slip
plane'' for the diamond (111).  One could naively expect that the strength of
the WWW sample, consisting of four-fold atoms only, could perhaps be higher
than that of diamond, due to the lack of such easy slip planes. On the
contrary, the calculation reveals that the strength of diamond for tensile
load along its easy slip direction, (111), is about twice that of the
isotropic WWW. Apparently, the lack of easy slip planes in WWW is compensated
by its somehow distorted bonds and the lack of order beyond the
first-nearest-neighbor distance.

We applied the same methodology to a na-C sample where crystalline regions
having radii of about 1.2 nm are embedded in ta-C with a volume fraction of
about 30\%. The stress-strain curve for this sample follows exactly that of
the embedding ta-C. The crystalline phase remains unaffected by the external
load, which is almost completely taken by the surrounding amorphous matrix.
Therefore, the response of na-C to external load beyond the elastic regime is
identical to the response of the embedding matrix. As atoms in the amorphous
matrix form bonds that are always weaker than the bonds in the crystal, the
system prefers to stretch or bend these bonds and keep the strong diamond
bonds untouched. By performing an atom-by-atom analysis of the deformation, we
can probe the four-fold atoms of the amorphous atoms as the ones more
extensively deformed when the material experiences large load \cite{fyta06}.

\section{Conclusions}
\label{sec:conclusions}

We examined theoretically the structure, elastic and inelastic response to
load of several carbon-based materials, including diamond, amorphous carbon
(a-C) ``amorphous diamond'' (WWW) and nanocomposite amorphous carbon (na-C).
These materials are formed by covalently bonded four-fold and three-fold atoms
and are characterized by their average coordination number ($z$). In a-C,
three-, four- and five-member rings are formed, their number increasing with
decreasing $z$. Most of such small rings contain four-fold atoms, while larger
rings contain also three-fold atoms. The bulk modulus of all these
carbon-based materials seems to follow a universal functional dependence of
the density. All elastic constants were also found to increase with increasing
density.

The strength of a-C was found to increase in roughly a linear manner, with
increasing concentration of four-fold atoms. High-density sample exhibited a
brittle behavior, analogous to that of diamond. The strongest a-C sample we
considered was the ``amorphous diamond'' WWW sample; this has a maximum stress
about half that of diamond. The response of na-C to external load is
essentially identical to the response of the embedding a-C matrix.

\section{Acknowledgment}

This work is supported by the Ministry of National Education and
Religious Affairs of Greece through the action ``E$\Pi$EAEK''
(program ``$\Pi\Upsilon\Theta$A$\Gamma$OPA$\Sigma$''.)



\end{document}